\documentclass [twocolumn,aps,prl,superscriptaddress]{revtex4-2}

\usepackage{amsmath}
\usepackage{graphicx}
\usepackage{MnSymbol}

\begin{document}

\title{Heat transfer in granular media with weakly interacting particles}
%https://www.overleaf.com/project/62ca1a56e5f26c8b9dbb5345
% Version Jens 10.7.2022 assuming T=200 K for figs 9, 10, 11 and T = 200 K we get  a(T ) ≈ 1.0 × 10−12 W~K and some addititions and corrections. 
\author{B.N.J. Persson}
\affiliation{Peter Gr\"unberg Institute (PGI-1), Forschungszentrum J\"ulich, 52425, J\"ulich, Germany, EU}
\affiliation{MultiscaleConsulting, Wolfshovener str 2, 52418 J\"ulich, Germany, EU}
\author{J. Biele}
\affiliation{German Aerospace Center, DLR, 51147 K\"oln, Germany, EU}

\begin{abstract}
{\bf Abstract}: 
We study the heat transfer in weakly interacting particle systems in vacuum.
The particles have surface roughness with self-affine fractal properties, as expected for mineral
particles produced by fracture, e.g., by crunching brittle materials in a mortar.
We show that the propagating electromagnetic (EM) waves and the evanescent EM-waves, which occur outside
of all solids, give the dominant heat transfer for large and small particles, respectively, while
the phononic contribution from the area of real contact is negligible.
As an application we discuss the heat transfer in rubble pile asteroids.

\end{abstract}

\maketitle

\setcounter{page}{1}
\pagenumbering{arabic}

%\pagestyle{empty}

%%%%%%%%%%%%%% main text %%%%%%%%%%%%%%%%
%\begin{multicols}{2}

%%%%%%%%%%%%%% main text %%%%%%%%%%%%%%%%

{\bf 1 Introduction}

Granular materials can be described as homogeneous media in the continuum approximation, on length scales
much larger then the particle sizes (diameter $2R$).
The effective thermal conductivity $K$ is an important 
property of granular materials which may be very different
from the thermal conductivity $K_0$ of a solid block made 
from the same material\cite{T1,T2}. 
Thus $K$ depends on not only $K_0$ but on the size and shape of the particles, on the nature of the 
particle contact regions,
and on the fraction of the total volume occupied by the particles, the so called filling factor (= 1 - macroporosity).
It also depends on the environment such as temperature, gas pressure and humidity.  

In this study we will assume that $K<<K_0$ as is typically the case
for granular solids in vacuum where there is no gas or fluid 
which could facilitate the heat transfer
between the particles. In this case, if all the particles are small enough, 
the temperature in each particle may be nearly constant
but the temperature change slightly from one particle to a nearby particle.
The effective thermal conductivity is determined by the heat transfer between the particles
and from dimensional arguments one expect 
$K \approx G/R$, where $G$ is the thermal contact conductance relating the
heat transfer rate between two particles to the temperature difference $\dot Q = G(T_0-T_1)$.

Most studies of the heat transfer in granular materials have assumed spherical particles without
surface roughness. However, all solids have surface roughness\cite{fractal} which affect 
all contact mechanical\cite{Mueser}, electrical\cite{elec} and thermal properties\cite{Heat1}.
In this paper we will study the influence of surface roughness on the heat transfer between particles
in granular media. A similar study but using a very different formalism, and with less realistic
types of surface roughness, was presented by Kr\"uger et al (see Ref. \cite{Kard}).

As an application we will consider the conduction of heat in asteroids\cite{Sakatani17,Sakatani18} or in regolith, the loose material covering many surfaces of solar system bodies.
% JB regolith is much more universal and an important keyword here!
Many asteroids consist of
weakly interacting particles of different sizes (rubble pile asteroids). 
The effective thermal conductivity of a thin surface layer of
asteroids can be obtained from the measured heat radiation, and are
typically found to be a factor 
$\sim 0.01$ times smaller than expected from a solid block made from the same material\cite{value}.
We will show that the heat conduction is mainly due to the radiative (for large particles) 
and evanescent (for small particles) electromagnetic field, while the
contribution from the area of real contact is negligible for weakly
interacting particles. Our result constitutes a totally new, 
revised concept of what is conventionally called the ``solid'' 
thermal conductivity in granular matter studies.

\begin{figure} [tbp]
\includegraphics [width=0.35\textwidth,angle=0]{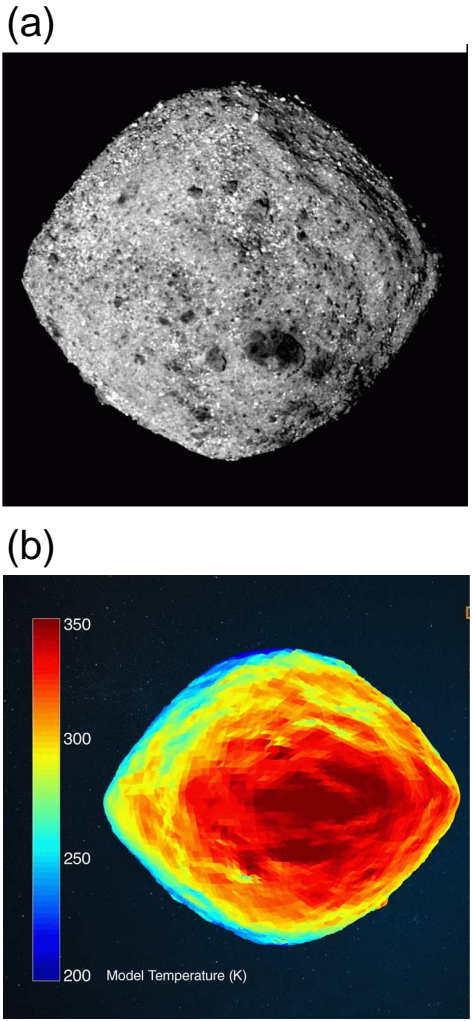}
\caption{
(a) Optical image of the asteroid Bennu.
(b) The temperature distribution obtained from Bennu thermal emission.
Credit: NASA/Goddard/University of Arizona 
}
\label{Temp2.ps}
\end{figure}

\vskip 0.3cm
{\bf 2 Heat transfer in asteroids}

All asteroids rotate and their surfaces are exposed to energy flux from the sun.
Thus the asteroid surface temperature will vary in space and time 
and this can be studied using infrared detectors. As an example, Fig. \ref{Temp2.ps}(a) shows an optical
image of the asteroid Benny and Fig. \ref{Temp2.ps}(b) the temperature distribution 
obtained from thermal emission. 

On the average an asteroid must emit as much heat radiation as it absorb from the sun.
Using this it is trivial to estimate the (average) asteroid surface temperature $T$ which is (in equilibrium) also the nearly constant interior temperature. The total sun
radiation power is $\dot Q \approx 4 \times 10^{26} \ {\rm W}$. Most asteroids (radius $r_{\rm a}$)
are at distances $r_0\approx (3-5)\times 10^{11} \ {\rm m}$ from the sun. If we assume that all the radiation (photons)
from the sun hitting an asteroid is absorbed by the asteroid then the absorbed power
$\dot q = \dot Q \pi r_{\rm a}^2 /(4 \pi r_0^2)$. The thermal radiation from the asteroids is the so-called ``fast-rotator model'' is
$\sigma T^4 4 \pi r_{\rm a}^2$, where $\sigma \approx 5.67\times 10^{-8} \ {\rm W/(m^2 K^4)}$ is the 
Stefan-Boltzmann constant.
Thus $\sigma T^4  4 \pi r_{\rm a}^2 = \dot Q \pi r_{\rm a}^2 /(4 \pi r_0^2)$ or
$$T= \left ({\dot Q \over 16 \pi \sigma r_0^2}\right )^{1/4},$$
which gives $T\approx 200 \ {\rm K}$. The asteroids in the asteroid belt between
Mars and Jupiter was formed at the same time as our solar system and is
hence $\approx 4.6 \times 10^9$ years old. In Ref. \cite{Biele} we showed that for a solid particle located in
vacuum for $\approx 4.6 \times 10^9$ years
at the temperature $T\approx 200 \ {\rm K}$ all molecules bound to the particles
by less than $\approx 1 \ {\rm eV}$ will have desorbed. Hence we can assume that no mobile
molecules exist which could form capillary bridges between the particles in asteroids.

The thermal emission of asteroids contains many important clues about their
physical properties and the study of asteroid thermal emission (often referred
to as thermal radiometry), together with optical brightness observations, 
is the primary source of known diameters and albedos (the measure of the diffuse 
reflection of solar radiation out of the total solar radiation),
and the only established remote-sensing means of determining the crucial thermal inertia defined below.
The principle of thermal radiometry is simple: Asteroids are heated up by
absorption of sunlight, the absorbed energy is radiated off as thermal emission. 
This leads to a characteristic $T(t,{\bf x})$ curve for each point 
${\bf x}$ on the surface, or for the disc-averaged infrared flux for non-resolved observations. 

When a solid is exposed to a photon energy flux $J(t)$ the temperature 
in the body will change with time. If the lateral variation of the energy 
flux $J$ is slow, and the time variation fast, the lateral
(parallel to the surface of the solid) heat diffusion can be neglected, and only heat diffusion normal to the
surface (coordinate $x$) need to be considered. In this case the surface temperature depends only on 
the thermal inertia defined as $\Gamma = (\rho C K)^{1/2}$,
where $\rho$ is the bulk mass density, $C$ the heat capacity and $K$ the thermal conductivity. 
This is easily seen for the simplest case
where the incident energy flux $J(t)$ oscillate harmonically in time. Writing
$$J(t)=J_0 {\rm e}^{-i\omega t}$$
$$T(x,t)=T_0 {\rm e}^{-i\omega t-\gamma x}+T_1$$
the heat diffusion equation 
$$\rho C {\partial T \over \partial t} = K {\partial^2 T\over \partial x^2}$$
gives
$$\rho C (-i\omega ) = K \gamma^2$$
so that
$$\gamma = \left (-i\omega \rho C K \right )^{1/2} K^{-1} $$
The boundary condition for $x=0$:
$$-K{\partial T\over \partial x} = J$$
gives
$$K\gamma T_0 = J_0$$
or
$$T_0 = {J_0 \over (-i\omega )^{1/2} \Gamma }$$
Thus the surface temperature depends only on the incident heat flux (amplitude $J_0$ and frequency $\omega$) 
and on the thermal inertia $\Gamma$. During one oscillation in the energy flux (period $2 \pi /\omega$)
the temperature penetrate a distance 
(the skin depth) $l \approx 1/{\rm Re}\gamma \approx K/(\Gamma \surd \omega )$ into the
solid.

Asteroids or regolith consist of solid particles of different sizes (e.g., \cite{Grott20}) and the results presented above are only
valid on length scales where the solid can be considered as effectively homogeneous, which would be length
scales larger then the particle diameter for particles with the same size.
Nevertheless, temperature map of the asteroids provide insight into their surface properties.
Thus low thermal inertia is typically associated with layers of dust, 
while high thermal inertia may indicate rocks on the surface.

\begin{figure} [tbp]
\includegraphics [width=0.35\textwidth,angle=0]{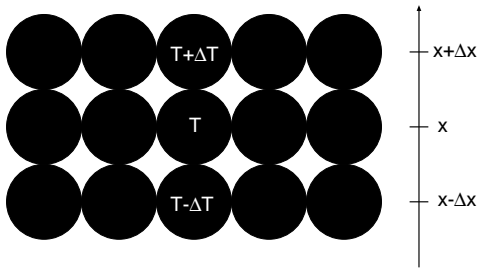}
\caption{
Heat flow in a simple cubic lattice of spherical particles (radius $R$). We assume the temperature
depends only on one coordinate direction denoted by $x$. Random packings of various porositys are considered in appendix C. 
}
\label{HeatFlow.eps}
\end{figure}

\begin{figure} [tbp]
\includegraphics [width=0.30\textwidth,angle=0]{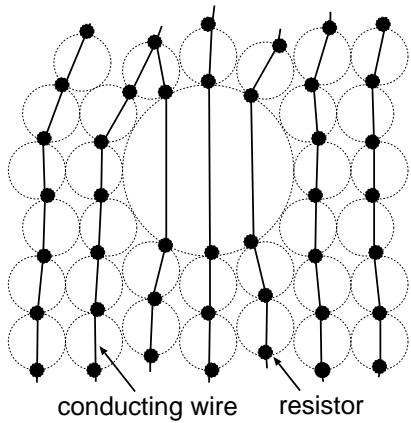}
\caption{
Heat flow in a collection of spherical particles (radius $R$) containing a particle with bigger radius. 
The system can be replaced by an electric analogy consisting of perfect conduction wires (line segments)
with resistors (black dots). The big particles will reduce the flow resistance of the system by effectively reducing the number
of resistors along some wires while keeping the number of wires (or flow channels) unchanged. 
Thus embedding big particles in a matrix of smaller particles may increasse the thermal conductivity, as indeed observed
experimentally (see Fig. 8 in Ref. \cite{T1}).
}
\label{Conductor.eps}
\end{figure}

\vskip 0.3cm
{\bf 3 Heat diffusion}

We assume that the heat transfer between the particles is so slow that the temperature in each particle is approximately
constant in space. We assume for simplicity that all the particles are spherical with equal radius $R$ and forming a 
simple cubic lattice,
and that the temperature depends only on the $x$-coordinate.
Consider a particle at at position $x$ at the temperature $T$ (see Fig. \ref{HeatFlow.eps}). 
The heat transfer rate from the particle at $x-\Delta x$ into the
particle at $x$ is $G [T(x-\Delta x)-T(x)]$ and the heat flow rate out of the particle at $x$ to the particle at $x+\Delta x$ is
$G [T(x)-T(x+\Delta x)]$, where $G$ is the (thermal) contact conductance. The net heat flow into the particle is
$$\dot Q = G [T(x-\Delta x)-T(x)] - G [T(x)-T(x+\Delta x)]$$
$$ \approx G {\partial^2 T\over \partial x^2} \Delta x^2$$
This will give rise to a change in the temperature in the particle determined by
$$\rho C V {\partial T \over \partial t} = \dot Q$$
where $\rho$ is the particle mass density, $C$ the heat capacity per unit mass and $V=4 \pi R^3/3$ the volume of the particle.
Thus we get
$$\rho C V {\partial T \over \partial t} \approx G {\partial^2 T\over \partial x^2} \Delta x^2$$
Using that $\Delta x = 2 R$ we get the heat diffusion equation
$$\rho C {\partial T \over \partial t} = K {\partial^2 T\over \partial x^2}$$
where the effective heat conductivity 
$$K={3\over \pi} {G\over R} . \eqno(1) $$ 
The prefactor $\gamma =3/\pi$ was derived for a simple cubic arrangements of the particles.
Other arrangements of the spheres result in similar expressions for $K$, 
but with different prefactor $\gamma$ (of order unity) (see Appendix A). 
We can also write
$${\partial T \over \partial t} = D {\partial^2 T\over \partial x^2}$$
where the heat diffusivity
$$D= {3\over \pi} {G\over \rho C R} .$$

Note that the thermal diffusivity $D$ and the effective heat conductivity $K$ both have a factor $1/R$.
Thus, since for small particles with very rough surfaces $G$ is nearly independent of the radius of the
particles (typically, $G\sim R^{0.2}$, 
see Sec. 3), 
% exponent varies with s (and H)
both $D$ and $K$ will increase as the size of the particles decrease. This counter-intuitive result
is easy to understand from the electric analogy shown in Fig. \ref{Conductor.eps}: 
decreasing the radius of the particles increases the number of contact points along
the heat flux lines proportional to $1/R$ but at the same time the number of flux lines increases as $1/R^2$.
As a result the heat resistivity will decrease as $R^2/R = R$ and the heat conductivity increase as $1/R$.
For big particles the (radiative) black-body radiation will give a contribution to $G$ proportional to
$R^2$ and in this case the effective heat conductivity $K$ will increase with the particle size as $K \sim R$. 
For a system consisting of big particles surrounded by a matrix
of small particles the 
effective heat conductivity may be determined by the small particles and the volume fraction they occupy
in the mixture (see Fig. \ref{Conductor.eps} for an electric analogy, and also Sec. 4).
We note that the effective conductivity of a system with randomly distributed particles of different sizes 
and thermal conductivity could be studied using an effective medium approach.

\begin{figure} [tbp]
\includegraphics [width=0.20\textwidth,angle=0]{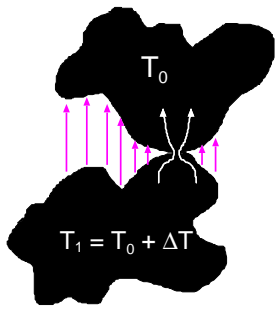}
\caption{
The heat transfer between two particles in vacuum can occur via lattice vibrations (phonons) 
in the area of real contact (white arrows)
or by electromagnetic radiation (photons) in the non-contact area (pink arrows).
}
\label{ParticleTemp.eps}
\end{figure}

\vskip 0.3cm
{\bf 4 The thermal contact conductance $G$}

Natural mineral particles, e.g., stone powder produced by crunching (involving fracture), are not perfect spherical
but have very rough surfaces (see Fig. \ref{ParticleTemp.eps}). This will drastically influence the 
heat contact conductance. In the most general case the heat transfer between two particles can occur via several different processes:

(a) Contribution from the area of real contact. For insulators this corresponds to heat transfer via phonons (lattice vibrations) 
which can propagate from one solid to the other via the area of real contact.

(b) Heat radiation. Here one must in general consider both the propagating electromagnetic (EM) waves, which corresponds to the
normal black-body radiation, and the evanescent EM-waves, which decay exponentially with the distance from the surfaces of the
solids. The latter will dominate the heat transfer at short surface separation and will be very important for small particles.

(c) Heat transfer can occur in the surrounding atmosphere by heat conduction (or convection) in the gas.

(d) In the normal atmosphere fluid capillary bridges may form and heat diffusion in the fluid bridges 
will contribute to the heat transfer. 

Here we are interested in the heat flow in asteroids and in these case there is no atmospheric gas and no capillary bridges
so only processes (a) and (b) are relevant.

\vskip 0.1cm
{\bf  Area of real contact contribution to $G$}

The thermal resistance of a contact is usually assumed to be due mainly to the constriction resistance.
However, this assumes that the material at the interface interact as strongly as in the bulk. This may be the case in many
practical (engineering) 
applications where the material in the contact regions are plastically deformed and where (for metals) cold welded regions form.
However, here we are interested in the contact between particles which interact very weakly. In this case most of the contact resistance
may be due to the weak coupling between the solids at the interface. We will now show that this is the case for the interaction
between particles in asteroids.

For an atomic-sized contact the interfacial heat conductance at high enough 
temperature (see Appendix B and Ref. \cite{Ryberg,Heat1,Heat2,rev1}):
$$G_{\rm a} \approx k_{\rm B} \eta \eqno(2)$$
where $k_{\rm B}$ is the Boltzmann constant and $\eta$ a damping due to phonon emission given by
$$\eta \approx {k^2 \mu \over \rho m {\rm c_{\rm T}^3}}\eqno(3)$$
where $c_{\rm T}$ is a transverse sound velocity, $\rho$ the mass density of the solids and $m$ the atomic mass. 
The parameter $\mu$ is given by (B6) (see also Ref. \cite{Ryberg}) 
and depends only on the the Poisson ratio $\nu$, or on the ratio 
$c_{\rm T}/c_{\rm L}$ between the transverse and longitudinal sound velocity.
For $c_{\rm T}/c_{\rm L} = 1/2$ one gets $\mu \approx 0.13$.
The heat conductance $G_{\rm a}$ for any temperature is obtained by replacing $k_{\rm B}$ in (2)
with $C_V/3$ where $C_V$ is the heat capacity per atom. At high temperatures $C_V = 3 k_{\rm B}$ (Dulong-Petit law) 
and we recover (2). Note that the heat capacity can be written as an energy fluctuation term (see Eq. (B7)) so the
heat conductance (3) can be expressed as an energy fluctuation term times a damping (or inverse relaxation time) term $\eta$.

In (3) occurs the spring constant 
$k=U''(d_0)$ where $U(d)$ is the interaction potential between surface groups on the two solids and $d_0$ the equilibrium separation.
We assume a Lennard-Jones (LJ) interaction potential:
$$U(d) = \epsilon \left [\left ({d_0\over d}\right )^{12} - 2 \left ( {d_0\over d}\right )^{6}\right ] .$$ 
Expanding in $d-d_0$ to second order gives
$$U(d) \approx -\epsilon+36 { \epsilon \over d_0^2 } (d-d_0)^2$$
Thus $k=72 \epsilon /d_0^2$ and using the same LJ-parameters as in Ref. \cite{Biele},
$\epsilon = 3.3 \times 10^{-21} \ {\rm J}$ 
(or $0.02 \ {\rm eV}$) and $d_0 = 0.3 \ {\rm nm}$, gives $k \approx 2.6 \ {\rm N/m}$.
Using the average atomic mass (the mass of ${\rm SiO_2}$ divided by 3) 
$m \approx 3\times 10^{-26} \ {\rm kg}$, the silica mass density $\rho \approx 2600 \ {\rm kg/m^3}$
and the sound velocity $c_{\rm T} \approx 4000 \ {\rm m/s}$ 
we get $\eta \approx 10^{11} \ {\rm s}^{-1}$. Thus the 
heat conductance $G_{\rm a} = \eta k_{\rm B} \approx 10^{-12} \ {\rm W/K}$ and 
the contact resistance $R_{\rm a} = 1/G_{\rm a} \approx 10^{12} \ {\rm K/W}$. We note that this
thermal resistance is even larger then found for molecular junctions (where typically $R \approx 10^{11} \ {\rm K/W}$)
involving small molecules between two gold electrodes, but in these cases the molecules are chemically attached to the
metal electrodes\cite{Got1,Got2}. 

\begin{figure} [tbp]
\includegraphics [width=0.35\textwidth,angle=0]{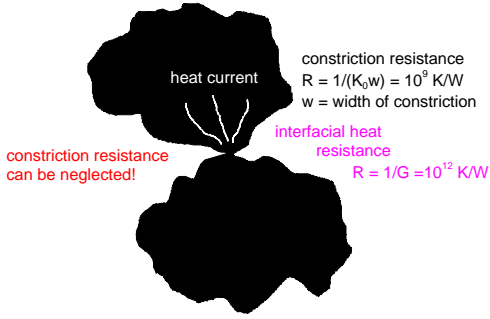}
\caption{
The (phononic) interfacial contact resistance and the constriction resistance act in series
and since the former is much larger the latter can be neglected.
}
\label{Negligible1.eps}
\end{figure}

The calculation above assumes implicitely that the bulk thermal conductivity is infinite.
For finite thermal conductivity there will be an additional contact constriction resistance.
We will show that the interfacial resistance is much larger than the constriction resistance 
$ 1 /(2 r_0  K_0)$ where $K_0 \approx 1 \ {\rm W/Km} $ is the silica thermal
conductivity and $2 r_0 \approx 1 \ {\rm nm}$ the diameter of the contact area. 
Thus $1/(2 r_0 K_0 ) \approx 10^{9} \ {\rm K/W}$.
Since the constriction resistance and the interfacial resistance $R_{\rm a}$ act in 
series they add together. Hence the
constriction resistance can be neglected and the contact conductance is 
determined by (2). This conclusion is summarized in Fig. \ref{Negligible1.eps}.

\vskip 0.1cm
{\bf Radiative contribution to $G$: theory}

The heat flux per unit area between two black-bodies with flat surfaces
(of area $A_0 \propto R^2$) separated by $d$ is given by the Stefan-Boltzmann law
$$J= {\pi^2 k_{\rm B}^4 \over 60 \hbar^3 c^2} \left (T_1^4-T_0^4\right )\eqno(4)$$
where $T_1$ and $T_0$ are the temperatures of solids ${\bf 1}$ and ${\bf 0}$, respectively,
$\hbar $ the reduced Planck's constant and $c$ the light velocity. 
Note that (4) is only valid if the surface separation $d$ is larger than the wavelength $\lambda$ of the
emitted radiation. Since $c k = \omega$ where the wavenumber $k=2 \pi /\lambda$ we get $\lambda = 2\pi c /\omega
=2 \pi c \hbar /\hbar \omega \approx 2 \pi c \hbar/k_{\rm B}T$ where we have used that a typical photon energy
$\hbar \omega$ is of order $k_{\rm B}T$. At $T=273 \ {\rm K}$ we get $d_T = c \hbar/k_{\rm B}T \approx 10 \ {\rm \mu m}$.

We assume $T_1-T_0 = \Delta T << T_0$ and expand $J$ to first order in $\Delta T$:
$$J \approx {\pi^2 k_{\rm B}^4 \over 60 \hbar^3 c^2} 4 T_0^3 \Delta T$$
Hence the radiative heat transfer coefficient
$$\alpha_{\rm r} \approx {2\pi^2 k_{\rm B}^4 \over 30 \hbar^3 c^2} T_0^3\eqno(5)$$ 
and the interfacial heat conductance $G_{\rm r} \approx \alpha_{\rm r} A_0$.

In this limiting case the heat transfer between the bodies is determined by
the propagating electromagnetic (EM) waves radiated by the bodies and does not depend on the separation
$d$ between the bodies. Electromagnetic waves (or photons) always exist outside any body due to thermal
or quantum fluctuations of the current density inside the body. The EM-field created by
the fluctuating current density exists also in the form of evanescent waves, which are damped exponentially
with the distance away from the surface of the body. For an isolated body, the evanescent waves do not give
a contribution to the energy radiation. However, for two solids separated by $d < d_{T}$, the heat transfer may
increase by many orders of magnitude due to the evanescent EM-waves; this is often referred to
as photon tunneling. 

For short separation between two solids with flat surfaces ($d << d_{T}$) the heat current due to the
evanescent EM-waves is given by\cite{rev1,Pol,Pen}
$$J = {4\over (2\pi)^3} \int_0^\infty d\omega \ \left (\Pi_0(\omega)-\Pi_1(\omega)\right )$$ 
$$\times \int d^2q \ e^{-2qd} {{\rm Im} R_0(\omega) {\rm Im} R_1(\omega) \over
|1-e^{-2qd} R_0(\omega)R_1(\omega) |^2}\eqno(6)$$
where
$$\Pi (\omega) = \hbar \omega \left (e^{\hbar \omega /k_{\rm B}T}-1\right )^{-1}$$
and
$$R(\omega) = {\epsilon (\omega) -1 \over \epsilon (\omega) + 1}$$
where $\epsilon (\omega)$ is the dielectric function, orientations average if anisotropic.
We assume again $T_1=T_0+\Delta T$ with $\Delta T << T_0$. Expanding (6) to linear order in $\Delta T$ gives $J=\alpha_{\rm e} \Delta T$ with
$$\alpha_{\rm e} = {k_{\rm B}\over \pi^2 d^2} \int_0^\infty d\omega \ \eta^2 {e^\eta \over (e^\eta -1)^2}$$ 
$$\times \int_0^\infty d\xi \xi \ e^{-2\xi} {{\rm Im} R_0(\omega) {\rm Im} R_1(\omega) \over
|1-e^{-2\xi} R_0(\omega)R_1(\omega) |^2}\eqno(7)$$
where $\eta = \hbar \omega /k_{\rm B}T$ and $\xi= q d$. Note that the second integral depends on $\omega$ so the first
integral involves also the result of the second integral.

\begin{figure} [tbp]
\includegraphics [width=0.47\textwidth,angle=0]{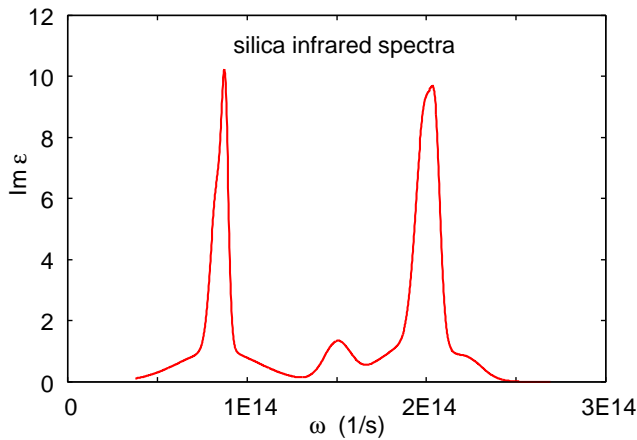}
\caption{
The imaginary part of the dielectric function of silica in the infrared region.
The two high peaks are due to two different optical phonons.
}
\label{Omega.2ImE.eps}
\end{figure}

\begin{figure} [tbp]
\includegraphics [width=0.47\textwidth,angle=0]{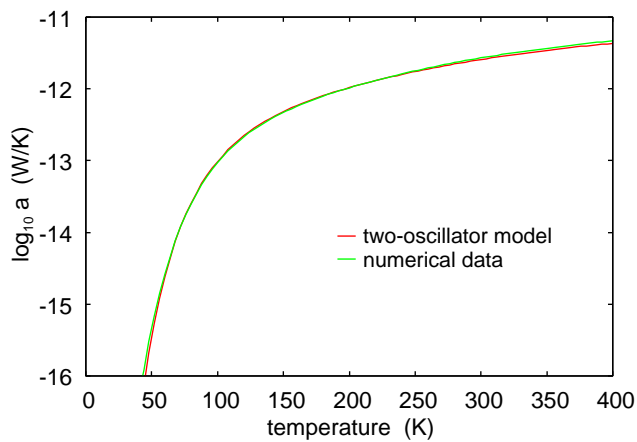}
\caption{
The $a(T)$ factor in the heat transfer coefficient $\alpha_{\rm e} = a/d^2$ as a function of temperature
for two silica surfaces (separation $d$). The red line is obtained using the two oscillator model (8)
while the green line is obtained using the numerical data for the dielectric function given in Ref. \cite{optical}.
}
\label{1Temp.2Alfadd.SiO2.eps}
\end{figure}

From (7)  it follows that the heat current scale as $1/d^2$ with the separation between the solid surfaces
and we write $\alpha_{\rm e} = a(T)/d^2$. 
For a spherical particle in contact with another particle (or a flat surface) the interfacial separation
$d \approx r^2/2R$ (where $1/R = 1/R_0+1/R_1$) where $r$ is the radial distance in the $xy$-plane from the
center of the contact region. In this case it is trivial to obtain the heat conductance
$$G_{\rm e} \approx \int d^2 r \ {a\over d^2} \approx 2 \pi R {a \over d_0}$$
where $d_0 \approx 0.3 \ {\rm nm}$ is the equilibrium separation between the surfaces at the contact point.

Consider now
two clean surfaces of (amorphous) silicon dioxide (SiO$_2$). The optical
properties of this material can be described using an oscillator
model\cite{optical}
$$\epsilon (\omega) = \epsilon_\infty + {a_1\over \omega_1^2 -\omega^2 -i\omega \gamma_1}
+{a_2\over \omega_2^2 -\omega^2 -i\omega \gamma_2}\eqno(8)$$ 
The frequency dependent term in this expression is due to optical phonons. 
The values for the parameters $\epsilon_\infty$, $(a_1,\omega_1,\gamma_1)$
and $(a_2,\omega_2,\gamma_2)$ are given in Ref. \cite{optical}.
In Fig. \ref{1Temp.2Alfadd.SiO2.eps} we show the temperature dependency of the $a(T)$ parameter
obtained using (8) (red line) and also as obtained using directly the measured $\epsilon (\omega)$ (green line).  
For $T=200 \ {\rm K}$ we get  
$a(T) \approx 1.0\times 10^{-12} \ {\rm W/K}$ and we use this value below when calculating the contact conductance $G$.
In Appendix C we show the $a(T)$ function for another mineral (olivine) of interest in applications to asteroids.
 
In the present case the heat transfer is associated with thermally excited optical (surface) phonons. That is, the electric field of a
thermally excited optical phonon in one solid excites an optical phonon in the other solid, leading to energy
transfer. The excitation transfer occur in both directions but if one solid is hotter than the other,
there will be a net transfer of energy from the hotter to the colder solid. 
%For metals, low-energy excited electron-hole pairs 
%will also contribute to the energy transfer. However, if the metals are covered with metal oxide layers, and if the separation
%between the solids is smaller than the oxide layer thickness, the energy transfer may again be due mainly
%to the optical phonons of the oxide, and the magnitude of the heat current will be similar to what we calculated
%above for (amorphous) silicon dioxide.

\begin{figure} [tbp]
\includegraphics [width=0.45\textwidth,angle=0]{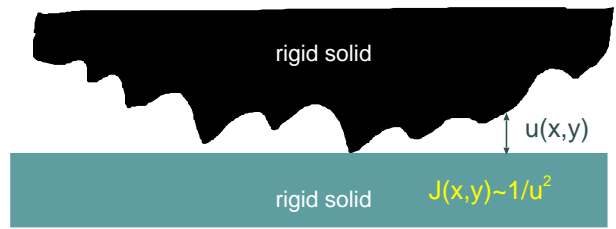}
\caption{
The heat current between the two surfaces is assumed to depend on the separation as $J(x,y) \sim u^{-2} (x,y)$.
}
\label{HeatPicJ.eps}
\end{figure}

\begin{figure} [tbp]
\includegraphics [width=0.45\textwidth,angle=0]{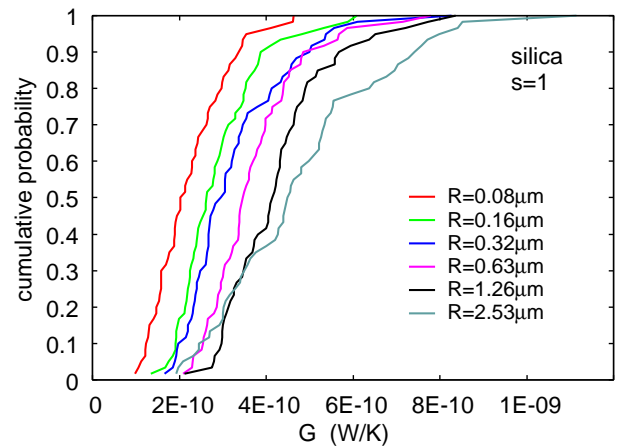}
\caption{
The cumulative probability for the evanescent $EM$ contribution to the contact conductance $G$. 
The probability distributions are obtained
from 60 simulations for each particle radius. The 60 simulations use
60 different realizations of the particle surfaces topography but with the same power spectra.
The calculations are for the granite surface (scaling factor s=1, Hurst exponent H=1) at 200 K.
}
\label{1G.2Cummulative.s=1.new.eps}
\end{figure}

\vskip 0.1cm
{\bf Radiative contribution to $G$: numerical results}

We have calculated the heat contact conductance using the same approach as used to study the adhesion between
particles with random surface roughness in Ref. \cite{Biele}.
No two natural stone particle have the same surface roughness, and the heat transfer between two particles
will depend on the particles used. To take this into account we have generated particles
(with linear size $L=2R$) with
different random surface roughness but with the same surface roughness power spectrum.
That is, we use different realizations of the particle surface roughness
but with the same statistical properties. For each particle size we have generated
60 particles using different set of random numbers. The surface roughness
was generated as described in Appendix A of Ref.  \cite{Rev} by adding plane waves with random phases
$\phi_{\bf q}$ and with the amplitudes determined by the power spectrum:
$$h({\bf x}) = \sum_{\rm q} B_{\bf q} e^{i ({\bf q} \cdot {\bf x} + \phi_{\bf q})}$$
where $B_{\bf q} = (2\pi /L)  [C({\bf q})]^{1/2}$. We assume isotropic roughness so $B_{\bf q}$ and $C({\bf q})$ only depend on the
magnitude of the wavevector ${\bf q}$. 

We have used nominally spherical particles with 6 different radii, where the radius increasing in steps of a factor of $2$ from
$R=78 \ {\rm nm}$ to $R=2.53 \ {\rm \mu m}$. The longest wavelength roughness which can occur on a particle with radius $R$
is $\lambda \approx 2R$ so when producing the roughness on a particle we only include the part of the power spectrum between
$q_0 < q < q_1$ where $q_0 = \pi /R$ and where $q_1$ is a short distance cut-off corresponding to atomic dimension
(we use $q_1 = 1.4\times 10^{10} \ {\rm m^{-1}}$). We will refer to these particles 
as granite particles because the power spectra used are linear extrapolation
to larger wavenumber of the measured granite power spectrum. For more details about the numerical procedure see Ref. \cite{Biele}.

We will now present numerical results for the heat conductance of granite particles.
We will also consider particles with the same sizes as above but with
larger and smaller surface roughness, obtained by scaling the
height $h(x,y)$ for the granite particles with scaling factors $s=0$ (smooth surface), $0.1$, $0.3$, $1$ and $2$.
Note that scaling $h(x,y)$ by a factor of $s$ will scale the power spectrum with a factor of $s^2$ but it will not change the
slope of the $C(q)$ relation on the log-log scale so the Hurst exponent (and the fractal dimension) is unchanged.

We assume that the heat current depends on the surface separation $u(x,y)$ as 
$J(x,y) \sim u^{-2} (x,y)$. This holds accurately only in the small slope approximation.
The heat conductance $G_{\rm e}$ is obtained by integration $\alpha_{\rm e} = a(T)/u^2(x,y)$ over the surface area
(see Fig. \ref{HeatPicJ.eps}).

Fig. \ref{1G.2Cummulative.s=1.new.eps}
shows the cumulative probability for the conductance $G$ for all the particles with different radius. 
The probability distributions are obtained by using for each particle size
60 different surface roughness realizations with the same power spectra.
The calculations are for the granite surface (scaling factor $s=1$).
Note that there is a slight increase in $G$ with increasing particle radius.
For the Van der Waals interaction between the particles the different particle radius
gave nearly the same cumulative probability distribution  i.e.,
the pull-off force, and the  statistical fluctuations in the pull-off force, where
nearly the same for all the particles. The reason for why $G$ exhibit a stronger (but still very weak)
dependency on the particle radius is that while $G$ depends (for flat parallel surfaces) 
on the interfacial separation $d$ as $1/d^2$ while the Van der Waals interaction decay faster as $1/d^3$; see, e.g., \cite{Krueger13}.

\begin{figure} [tbp]
\includegraphics [width=0.47\textwidth,angle=0]{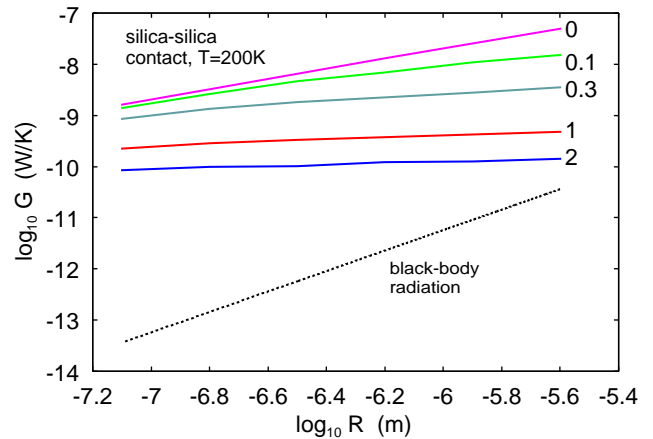}
\caption{
The heat contact conductance as a function of the particle radius (log-log scale) at $T=200 \ {\rm K}$, H=1, parameter s. The solid
lines are assuming the evanescent-wave electromagnetic coupling between the particles.
The dashed line is the result assuming propagating-wave electromagnetic coupling
(i.e. black body radiation) but this result is only for illustrative purpose because the Stefan-Boltzmann
law is not valid for the small surface separation occurring with particles as small as studied here.
The different solid lines are for particles where the surface roughness of the granite particle
is scaled with different factors s between  0 (smooth surface) and 2.
}
\label{1logR.2logAlpha.eps}
\end{figure}

\begin{figure} [tbp]
\includegraphics [width=0.47\textwidth,angle=0]{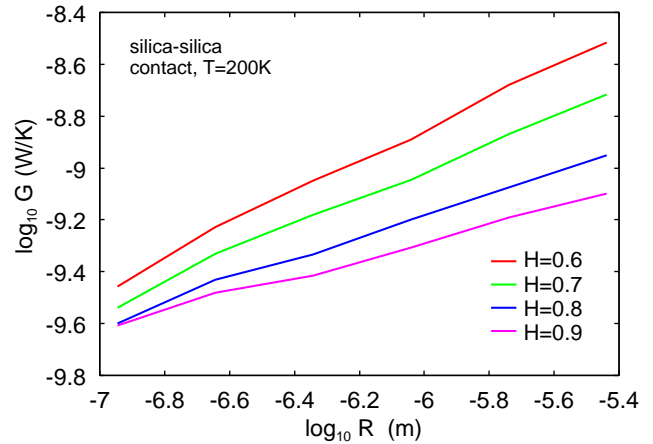}
\caption{
The heat  contact conductance as a function of the particle radius (log-log scale) at $T=200 \ {\rm K}$
for particles with self-affine fractal surface roughness with different Hurst exponent $H=0.6$, 0.7,
0.8 and 0.9; roughness amplitude s=1. All surfaces have the same root-mean-square slope when including the roughness on all
length scales (see Ref. \cite{Biele} for more details about the surface roughness).
}
\label{1logR.2logH.many.fractal.eps}
\end{figure}

\begin{figure} [tbp]
\includegraphics [width=0.47\textwidth,angle=0]{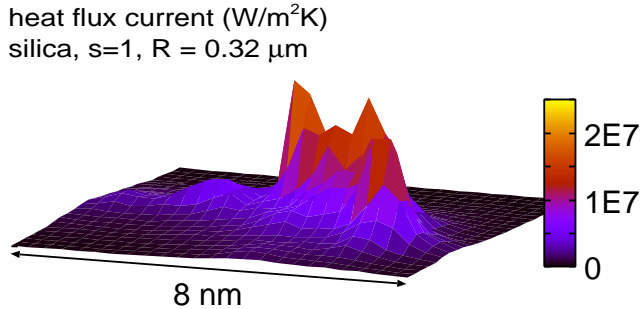}
\caption{
The heat current $\partial J(x,y,T)/\partial T$ close to the point where it is maximal.
For one realization of the surface roughness for $R=0.32 \ {\rm \mu m}$ and $T=273 ^\circ {\rm C}$. 
}
\label{topography.heat.flow.X3.eps}
\end{figure}

\begin{figure} [tbp]
\includegraphics [width=0.47\textwidth,angle=0]{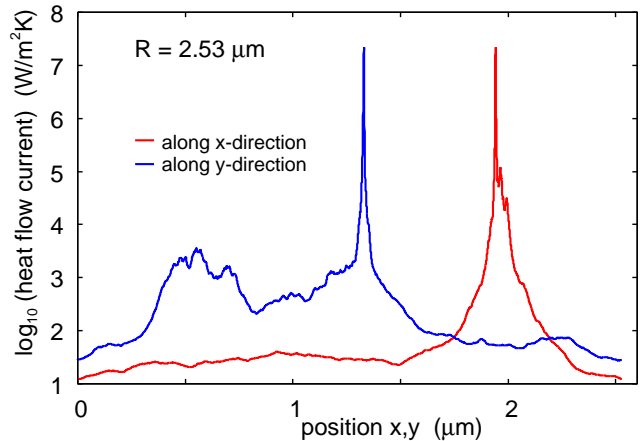}
\caption{
The heat current $\partial J(x,y,T)/\partial T$ along the $x$ and $y$-directions through 
the point where it is maximal. 
For one realization of the surface roughness for $R= 2.53 \ {\rm \mu m}$
and $T=273 ^\circ {\rm C}$. 
}
\label{1xy.2heatflow.R=2.53mum.eps}
\end{figure}

\begin{figure} [tbp]
\includegraphics [width=0.47\textwidth,angle=0]{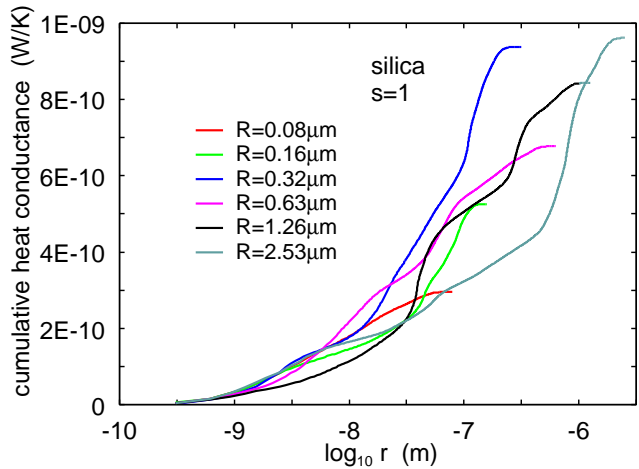}
\caption{
The cumulative heat conductance obtained by integration the heat current 
$\partial J(x,y,T)/\partial T$ over a circular region $|{\bf x} - {\bf x}_0|<r$ 
with the radius $r$, and centered at the point ${\bf x}_0$ where the heat current is maximal.
For one realization of the surface roughness for all the particles and $T=273 ^\circ {\rm C}$. 
}
\label{1logRadius.2cumulativeHeatConductance.s=1.eps}
\end{figure}

Fig. \ref{1logR.2logAlpha.eps} shows
the heat contact conductance as a function of the particle radius (log-log scale) at $T=273 \ {\rm K}$. The solid
lines are assuming the evanescent-wave electromagnetic (EM) coupling between the particles.
The dashed line is the result assuming propagating EM-wave coupling
(i.e. black body radiation) but this result is only for illustrative purpose because the Stefan-Boltzmann
law is not valid for the small surface separation occurring with particles as small as studied here.
The different solid lines are for particles where the surface roughness of the granite particle
is scaled with different factors between  0 (smooth surface) and 2.
Note that for particles with radius $R > 10 \ {\rm \mu m}$ the black-body radiation 
will dominate the heat transfer for granite particles but for particles with smooth surfaces ($s=0$) a
much larger particle radius is needed before the black body radiation dominates the heat transfer.
Note also that the surface separation $d$ must obey $d > d_T = c \hbar /k_{\rm B} T \approx 10 \ {\rm \mu m}$ (for
$T=273 \ {\rm K}$) in order for the Stefan-Boltzmann law to be valid.

Fig. \ref{1logR.2logH.many.fractal.eps} shows 
the heat conductance as a function of the particle radius (log-log scale) at $T=273 \ {\rm K}$
for particles with self-affine fractal surface roughness with different Hurst exponent $H=0.6$, 0.7,
0.8 and 0.9. All surfaces have the same root-mean-square slope when including the roughness on all
length scales (see Ref. \cite{Biele} for more details about the surface roughness).

\begin{figure} [tbp]
\includegraphics [width=0.35\textwidth,angle=0]{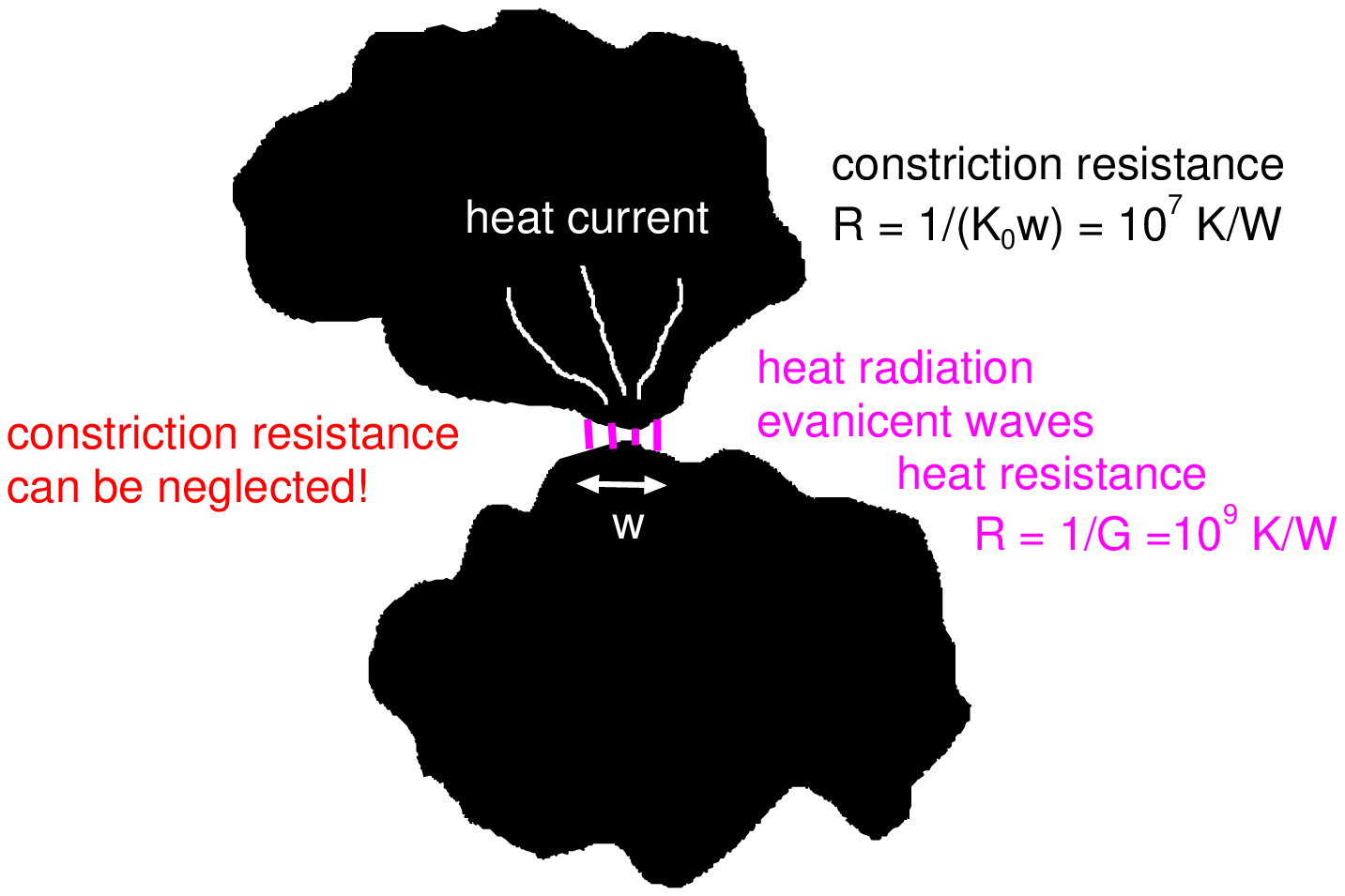}
\caption{
The EM (evanicent waves) contact resistance and the constriction resistance act in series
and since the former is much larger the latter can be neglected.
}
\label{Negligible.eps}
\end{figure}

Fig. \ref{topography.heat.flow.X3.eps} shows 
the spatial dependency of the heat current $\partial J(x,y,T)/\partial T$ close to the point where it is maximal.
The results is for one realization of the surface roughness for the particle with the radius 
$R=0.32 \ {\rm \mu m}$ and $T=273 ^\circ {\rm C}$. From the figure it may appear that most heat flow is localized to
a few nanometer sized region close to the point where the current in maximal. However, when plotted on a logarithmic
scale it is clear that the biggest contribution to the heat transfer occur from a much bigger surface region.
Thus, in Fig. \ref{1xy.2heatflow.R=2.53mum.eps}
we show the heat current $\partial J(x,y,T)/\partial T$ along the $x$ and $y$-directions through 
the point where it is maximal for a particle with the radius $R= 2.53 \ {\rm \mu m}$.
In Fig. \ref{1logRadius.2cumulativeHeatConductance.s=1.eps} we show for all the particles
the cumulative heat conductance obtained by integration the heat current 
$\partial J(x,y,T)/\partial T$ over a circular region $|{\bf x} - {\bf x}_0|<r$ 
with the radius $r$, and centered at the point ${\bf x}_0$ where the heat current is maximal.
Note that for the 3 biggest particles  
about half of the total heat conductance result from the heat flow within a circular
region with the radius $r\approx 0.1 \ {\rm \mu m}$. This is an important result because it implies that
the constriction contribution to the thermal resistance for the EM heat transfer can be neglected.
Thus if we assume $r_{\rm e} \approx 0.1 \ {\rm \mu m}$ we get the constriction resistance $1 /(2 K_0 r_{\rm e}) \approx
10^{7} \ {\rm K/W}$  
which is much smaller than the resistance $R_{\rm e} = 1/G_{\rm e} \approx 10^{10}  \ {\rm K/W}$
resulting from the evanescent EM waves. Since these resistance act in series we can neglect the
constriction contribution. Hence for very rough particles as produced by crunching mineral stone
one expect the contact conductance to be determined by the evanescent EM waves. 
This conclusion is summarized in Fig. \ref{Negligible.eps}.

In the calculation above we have assumed that both materials in the contact region between two particles
are identical. Many mineral particles consist of grains with (slightly) different chemical composition.
In that case the two reflection factors $R_0(\omega)$ and $R_1(\omega)$ in (7) will differ, which 
will reduce $a(T)$ as the heat transfer depends on the product ${\rm Im}R_0(\omega ){\rm Im}R_1(\omega)$, 
which depends on ${\rm Im}\epsilon_0 (\omega) {\rm Im}\epsilon_1 (\omega)$. 

For completeness me mention the "grain" thermal conductance, $G_p$,  which can be estimated by approximating the particle (material thermal conductivity $K_0(T)$) by a cube with equal volume, leading to
${{G}_{p}}={{\left( \frac{4\pi }{3} \right)}^{1/3}}R\:{{K}_{0}}$. 
$G_p$ in W/K is of the order of the grain diameter in m, in practice, $10^{-6}$ to $10^{-2}$ W/K, so small that it can in almost all cases be neglected. 

Note that the classical constriction resistance is based on continuum theory and neglects the phonon interface (acoustic mismatch) resistance $1/G_{I/F}$,  which  exists even for strong coupling if the bonded lattices are even slightly different or have a different orientation (usually this is the case at grain boundaries). A typical value \cite{Jin16} is $10^{8} \ {\rm W/m^2/K}$ for the conductance per area, so e.g. for  $10 \ {\rm \mu m}$ grains, $G_{I/F} \approx 0.01 \ {\rm W/K}$.

\vskip 0.3cm
{\bf 5 Comparison with experiment: a proposal}

The theory presented above could be tested experimentally by measuring the heat conductivity
of granular materials under high vacuum condition. We suggest to use
crunched minerals sieved to obtain monodisperse particle size distribution. We suggest to use crunched pure minerals (crystalline preferred), sieved to obtain a monodisperse particle size distribution. The roughness power spectrum of the mineral particles shall be determined and their dielectric function in the mid-IR must be known. Thermal conductivity measurements should be done in vacuum, after baking out the samples to make sure the particles lost all their physisorbed  water. 
Caution ahould be exercised w.r.t. boundary conditions in all experiments, whether it be a line heat source, guarded hot plate, strip heat source, or whatever: the packing structure (and contact resistance) is always disturbed near a boundary or implanted wire/strip. 
Measurements should be performed for particle sizes from $\sim 1 \ {\rm \mu m}$ to at least  $\sim 500 \ {\rm \mu m}$  such that the transition from dominating near-field EM heat transfer to, additionally,  firmly geometrics optics radiative far-field conductivity can be observed.
The temperature range should be as large as possible, to confirm the temperature dependence of $a(T)$ and  to be able to separate the contact conductance from the radiative (far field) conductivity, which ought to be possible for very low (say 80 K) and very high (say 600 K) temperatures, where $d {\rm log}(a)/d {\rm log} T$ typically has slopes quite different from 3. Temperatures down to 25 K with e.g. crystalline corundum (synthetic sapphire) ${\rm Al}_2{\rm O}_3$ would be interesting too, since this crystal has a very large (200 W/K/m) bulk thermal conductivity peak at ~30 K, two orders of magnitude higher than at room temperature \cite{Berman52}; this should be clearly seen in the data if solid conduction were phonon-dominanted via strong contacts.
Various porosities from random loose to random close packing should be prepared by a dedicated tapping procedure and determined precisely. 
It would also be interesting to perform the same experiments in the normal atmosphere at different humidities to include capillary bridges and heat diffusion in the fluid and in the surrounding air\cite{Heat1}.
Additionally, lithostatic pressure could be applied to study the its effect on contact conduction. To analyze the latter experiments one would need to extend the study we present in this paper.

Interesting heat conductivity data was presented by Sakatani et al. \cite{Sakatani17}. 
They used spherical glass beads with radii ranging from $2.5 \ {\rm \mu m}$ to $427  \ {\rm \mu m}$.
The roughness of the glass beads was not studied in detail but SEM images showed non-random roughness.
The experiments was performed in vacuum, but liquid bridges 
may have formed while handling the granular material in humid lab air before the vacuum study.
Experiments have shown that under the influence of water or humidity
Si-O-Si bonds may form between two silica surfaces in the contact regions which could 
remain in the vacuum condition and reduce the interfacial contact
resistance\cite{bond,PRL1,PRL2,PRL3}. However, if the granular media is exposed to vibrations when in the vacuum
then these chemical bonds may be broken in which case the particle-particle interaction may be mainly 
of the Van der Waals type as assumed here. 
For these reason no quantitative comparison of the data of Sakatani et al. 
with the present theory is possible or meaningful.

\begin{figure} [tbp]
\includegraphics [width=0.47\textwidth,angle=0]{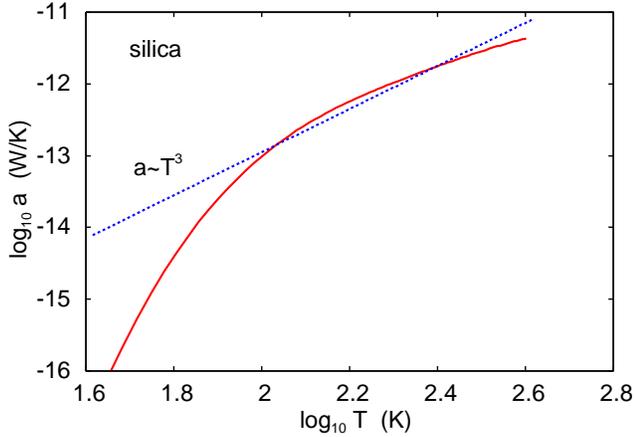}
\caption{
The $a(T)$ factor as a function of temperature 
(from Fig. \ref{1Temp.2Alfadd.SiO2.eps}) on a log-log scale (solid line).
The dotted line has a slope of 
$3$ corresponding to a temperature dependency $T^3$ expected from the black-body
heat transfer. 
}
\label{1logT.2loga.eps}
\end{figure}

\begin{figure} [tbp]
\includegraphics [width=0.47\textwidth,angle=0]{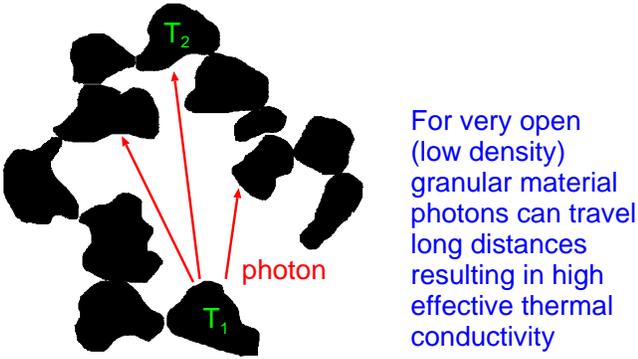}
\caption{
In very open (low filling factor) granular materials cavities much bigger than the particle
diameter can occur. In this case (Stefan-Boltzmann) heat radiation 
may result in large effective thermal conductivity.
}
\label{OPEN.eps}
\end{figure}

\vskip 0.3cm
{\bf 6 Discussion}

Most asteroids larger than $\sim 100 \ {\rm m}$ are rubble piles, i.e. consist of more than one solid objects, 
and are usually covered by a regolith layer (dust, broken rocks, and other related materials).
In fact all asteroids which have been studied with spacecraft consist of a wide distribution of fragments with
sizes ranging from about 100 meters to centimeter or less. The average temperature in the asteroids (typically
$\approx 200 \ {\rm K}$) is so high that on the time scale of billions of years
all loosely bound molecules have desorbed so one expect no mobile
adsorbed molecules which could form capillary bridges between the particles (or fragments). Since the
gravitational field is too weak to allow for an atmosphere the particles in asteroids are surrounded by vacuum.
Hence the heat transfer between the particles can only occur via the area of real contact (which are nanosized regions
as shown in Ref. \cite{Biele}) or via the fluctuating EM-field (propagating or evanescent) which occur  
around all solid objects. We have shown that because of the weak adhesion and gravitational force, the contribution to the
contact conductance from the area of real contact is negligible compared to the contribution from the EM-field. 

From studies of the surface temperature of asteroids one can deduce the thermal inertia 
$\Gamma = (\rho C K)^{1/2}$. Most asteroids have thermal inertia between $10$ and $1000 \ {\rm Jm^{-2}K^{-1}s^{-1/2}}$
and typically $\Gamma \approx 100 \ {\rm Jm^{-2}K^{-1}s^{-1/2}}$ (see Ref. \cite{value}). 
Assuming $\rho \approx 2\times 10^3 \ {\rm kg/m^3}$,
$C \approx 500 \ {\rm J/kg K}$  and $\Gamma \approx 100 \ {\rm Jm^{-2}K^{-1}s^{-1/2}}$ we get $K \approx 10^{-2} \ {\rm W/Km}$.
This is much smaller than the thermal conductivity of silicate rocks which is $\approx 1 - 10 \ {\rm W/Km}$. This proves
that most asteroids does not consist of single silica blocks but of weakly coupled fragments where most of the thermal
resistance comes from the thermal coupling between the fragments. 

Assume that all the particles in an asteroid would have equal size. In this case it is interesting to consider
two limiting cases where the heat transfer occur either via the radiative (black-body) EM-field, or via 
the evanescent EM-field. In the second case $G$ is nearly independent of the particle radius
and of order $G\approx 10^{-9} \ {\rm W/K}$ giving $R \approx G/K \approx 0.1 \ {\rm \mu m}$. In the first case
$G= \mu R^2$ where $\mu \approx 5.7 \ {\rm W/m^2K}$ for $T=200 \ {\rm K}$. 
This gives $R = K/\mu \approx 2 \ {\rm mm}$. 
In reality there is a wide distribution
of particle sizes and the heat transfer may involve both the radiative and the evanescent EM-wave interaction. However,
if the big particles (fragments) are surrounded by a matrix of much smaller (micrometer-sized) particles (dust), 
as suggested by other studies\cite{Biele}, then the radiative coupling my be less important than the evanescent contribution. 
In particular, unless the cavity regions between the particles are of size $d > d_T= c\hbar /k_BT\approx 10 \ {\rm \mu m}$ the 
the Stefan-Boltzmann law will not be valid. In this case the big particles will effectively act as regions
of infinite heat conductivity (because heat transfer in compact solids fast), 
and the resistance to the heat flow will occur mainly in the (low volume) matrix of small particles surrounding the bigger particles. 
Thus taking this effect into account, and assuming that the small particles have radius $R \approx 1-10 \ {\rm \mu m}$ as suggested by other
studies\cite{Biele}, may result in thermal inertia similar to what is observed.

Fig. \ref{1logT.2loga.eps} shows 
the $a(T)$ factor as a function of temperature (from Fig. \ref{1Temp.2Alfadd.SiO2.eps}) on a log-log scale (solid line).
The dotted line has a slope of $3$ corresponding to the temperature dependency $T^3$ expected from the black-body
heat transfer. Fig. \ref{1logT.2loga.eps} shows that the contribution to the heat transfer from the evanescent waves may 
increase faster or slower with temperature than the black body contribution, depending on the temperature region. 
%Experimentally, the thermal inertia of asteroids depends on the temperature typically as 
%$\Gamma \sim T^{2.7}$ (see Ref. \cite{value}).
%If $G\sim T^3$ and the heat capacity (as expected) $C\sim T$ one would get $\Gamma \sim T^2$. 
% Jens: this is disputed. Observations are inconclusive or just not accurate enough. Also C is not ~T, look at a Debye curve...

%However, asteroids do not contain the ``pure'' silica used in the calculation 
%of the contribution from the evanescent waves
%so from the temperature dependency of the measured thermal inertia one cannot decide which process dominate!
We note that in very open (high porosity) granular materials cavities much bigger than the particle
diameter can occur (see Fig. \ref{OPEN.eps}). In this case (Stefan-Boltzmann) heat radiation 
may result in large effective thermal conductivity. 
The (macro-)porosity $\phi$ [which equals (void volume)/(total volume)]
of rubble pile asteroids ranges from $\phi \approx 0.15$  (see Ref. \cite{Grott20}) to $\phi \approx 0.5$ (see Ref. \cite{pore}), and the very top layer of lunar regolith has a porosity $> 0.8$ (see \cite{HapkeMoon})  which may
result in some fraction of the emitted photons traveling one or several particle diameters.

In the asteroid research field it is usually assumed that the constriction 
resistance is a very important contribution to the thermal contact resistance.
We have shown that the constriction resistance is important
only when cold-welded (for metal) or sintered contact forms, or where strong (chemical) bonds form between the 
particles in the contact area. However, experiments have shown that 
the asteroid particles interact only with very weak forces and the contact resistance
is dominated by the interfacial thermal resistance (determined by (2)), and the constriction resistance can be neglected.

In asteroid research it is usually assumed that the thermal conductivity
$K_{\rm eff} = A+BT^3 $ where $A$ is assumed to be due to the constriction resistance 
and hence proportional to the bulk
thermal conductivity $K_0(T)$ which is only weakly temperature dependent. 
However, we find that the constriction resistance term can be neglected and that
$A$ may be due to the evanescent EM-waves $K_{\rm e} \propto a(T)$, where $a(T)$ typically has a 
strong $T$-dependence, where at intermediate temperatures it is roughly $\propto T^3$ . 
Thus, $K_{\rm e}$ may have a similar $T$-dependence as $K_{\rm r}$. 
The particle size-dependence of $K_{\rm e}$ is also fundamentally different, 
$K_{\rm e}\propto(1/R)R^n$, where $n \approx 0.2$,
while in the conventional theories the term $A$ is independent of $R$, 
or increases only for very small $R$ due to an (assumed) larger effect of van Der Waals adhesion deformation.  
This should be observable in data for small $R< \approx10 \mu\rm{m}$. 

Finally we note that the Hertz- or JKR-theory for the contact radii between two particles in the conventional 
theories is unrealistic. The particles have large roughness and are made from stiff materials 
and the JKR (and Hertz) theory is not valid. Instead the adhesive interaction is very 
weak as we already have shown\cite{Biele}. This also means that the theoretical dependence of 
$K$ on the external (e.g., lithostatic) pressure $P$, roughly $\propto P^{1/3}$ (plus a constant for $P=0$), 
is not due to the contact mechanics between smooth adhering spherical particles. 
What matters, is the reduction of the surface separations of the two rough particles in their 
contact zone upon an externally applied pressure, since this controls evanescent EM-wave heat transfer.

\vskip 0.3cm
{\bf Acknowledgments}

Part of this work was performed when Bo Persson was participating
in the UCSB, KITP program ``Emerging Regimes and Implications of Quantum and Thermal Fluctuational Electrodynamics''.
This research was supported in part by the National Science Foundation under Grant No. NSF PHY-1748958.

\vskip 0.3cm
{\bf Appendix A}

In Sec. 3 we considered heat diffusion in a simple cubic arrangement of spherical particles with radius $R$ and found
$K= \gamma G/R$ with $\gamma = 3/\pi$. The study assumed that 
the heat transfer occur only at, or in the vicinity of, the 
contact regions between the particles as expected for the phononic contribution $K_{\rm a}$ 
or the contribution from the evanescent EM waves $K_{\rm e}$. 
In Sec. 4 we also considered the radiative (black-body) contribution to the 
heat conductance $G_{\rm r} \approx \pi R^2 \sigma T^3$ so we expect
$K_{\rm r} = \beta \sigma R T^3$ where $\beta \approx 3$. Here we will give approximate 
expressions for $\gamma$ and $\beta$ for realistic distribution
of particles as derived by Arakawa et al\cite{Arakawa19} and by Ryan et al\cite{Ryan22}, respectively. 
We expect $\beta$ and $\gamma$ to be of order unity,
but they depend on the particle filling factor (or porosity), 
and reliable estimates of these factors are needed for comparison with experiments or observational data.  

\vskip0.2cm
{\bf Contact conductivity} $K$

For a general granular medium we have to modify equation (1) by a factor which depends
on the porosity, and on packing geometry (ordered or random; shape and friction of particles). 

Arakawa et al\cite{Arakawa19} have performed simulations of the heat 
transfer in granular media consisting of
micrometer-sized spherical particles with random packing
with the porosity ranging from 0.37 to 0.99. 
The simulated results where fitted to obtain the effective thermal conductivity 
$K=\gamma G/R$ with 
$$ \gamma =0.533 (1- \phi)^{1.99} C^{0.556} \eqno(A1)$$
where the average coordination number $C$ depends on the porosity $\phi$. 
For spherical particles 
$$ C(\phi )=2+9.38{{(1-\phi )}^{1.62}},$$ 
but for irregular particles $C$ is typically smaller than for spheres at the 
same porosity\cite{Wood20,Nagaashi21}. 

\begin{figure} [tbp]
\includegraphics [width=0.47\textwidth,angle=0]{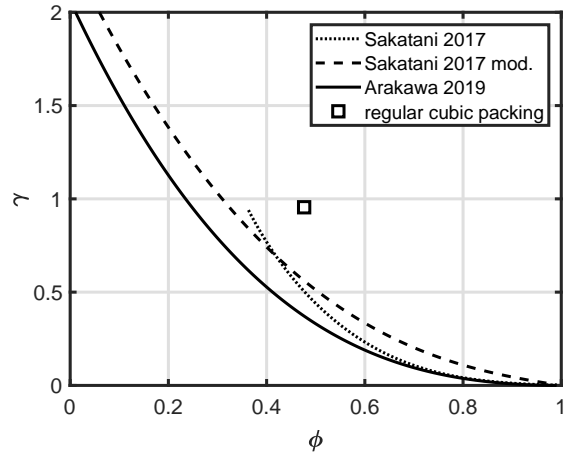}
\caption{
The geometrical factor $\gamma$ for heat diffusion via particle contacts, 
as a function of porosity (for random packings). 
After \cite{Arakawa19} (solid line) and \cite{Sakatani17} but with the coordination number model of \cite{Arakawa19}  (dashed line).  
The value for the regular cubic packing is indicated, too.
}
\label{alphavsphi.eps}
\end{figure}

For a simple cubic lattice $\phi = 1-\pi/6 \approx 0.476$
giving $C \approx 5.3$ and $\gamma \approx 0.31$ which is smaller than the factor $3/\pi \approx 0.95$
expected for a simple cubic lattice of spherical particles. The factor $\gamma$ is shown as a function of the
porosity in Fig. \ref{alphavsphi.eps} where the square symbol indicate $\gamma$ for the simple cubic lattice.

\vskip0.2cm
{\bf Radiative conductivity $K_{\rm r}$}

Ryan et al. \cite{Ryan22} have shown that for opaque particles in the 
geometrical optics regime $K_{\rm r} = \beta \sigma RT^3$ with
$$\beta =8 F(\phi) f(\phi). \eqno{(A2)}$$
The radiative exchange factor 
$$F=\epsilon \left[ a+b \left( {\phi \over 1-\phi}\right)^c \right], $$
where $\epsilon$ is the hemispherical total emissivity at thermal wavelengths; 
for most minerals $\epsilon\approx0.90$, but $\epsilon$ is weakly temperature-dependent.
For random particle packing $a = 0.739$, $b = 0.629$, and $c = 1.031$, and for
ordered packings  $a = 0.773$, $b = 0.419$, and $c = 1.180$.
The ``non-isothermal correction factor''
$$f=a_1  {\rm tan}^{-1} \left [ a_2 \left ({1-\phi \over \Lambda } \right )^{a_3} \right ] + a_4$$
where $a_1 = -0.500$, $a_2 = 1.351$, $a_3 = 0.741$, and $a_4 = 1.007$ and 
$${\Lambda}={K_0 (T) \over 8R\sigma\bar{T}^3}$$
For poly-disperse packings the diameter $D=2R$ in the equations above 
is (approximately) the Sauter mean diameter \cite{Kowal16}.

\begin{figure} [tbp]
\includegraphics [width=0.47\textwidth,angle=0]{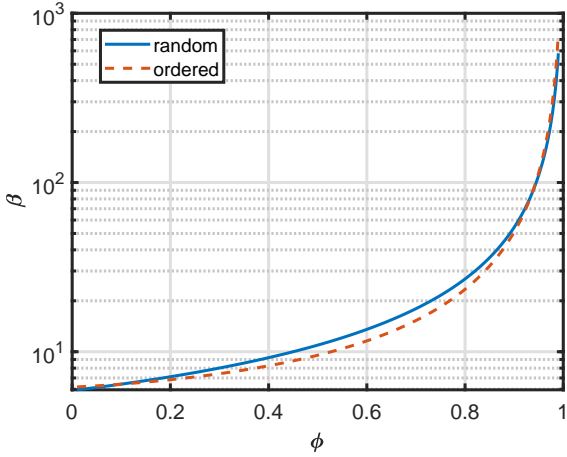}
\caption{
The geometrical factor $\beta$ for conventional radiative 
conductivity $K_{\rm r}=\beta \sigma R T^3$, as a function of porosity for emissivity $\epsilon=1$
in the limit $\Lambda \rightarrow \infty$. 
Adapted from \cite{Ryan22}.
}
\label{betavsphi.eps}
\end{figure}

For small enough spherical particles, or particles with high enough thermal conductivity
$K_0$, the temperature in the particles will be uniform
corresponding to $\Lambda \rightarrow \infty$. In this limit $f \approx 1$. 
For a simple cubic lattice of particles $\phi \approx 0.476$ and with $\epsilon = 1$ we get
$F \approx 1.15$ giving $\beta \approx 9.2$. Fig. \ref{betavsphi.eps} shows $\beta$ as a function of $\phi$
when $\Lambda = \infty$.

Equation (A2) is strictly valid only in the regime of geometrical optics, i.e., 
if the thermal wavelength (in meter) $2.9\times 10^{-3} \ T^{-1} $ (with $T$ in Kelvin) 
is larger than particle diameter $D$. 
The case  when the thermal wavelength is smaller than the particle size $D$ and/or the particles cannot be regarded as opaque to thermal radiation
was studied in Ref. \cite{Millan11,Arakawa17,Ito17,Grose19}. However, in this case
the contribution from the evanescent EM waves may be more important.

To facilitate the application of our new heat transfer model for the planetary sciences or quantitive comparisons with experimental data, we need correlation equations that represent $G_e$ as a function of $R$, $s$, $H$ and then the temperature-dependence ($T_0=200 {\rm K}$):

\[{{G}_{e}}={{\left. {{G}_{e}}(R,s,H) \right|}_{silica-silica,\,{{T}_{0}}}}\times \frac{{{a}_{\text{mineral0-mineral1}}}(T)}{{{a}_{\text{silica-silica}}}({{T}_{0}})}\]

As for  ${{\left. {{G}_{e}}(R,s,H) \right|}_{silica-silica,\,{{T}_{0}}}}$ it is possible to interpolate the results shown in figures 10, 11. Of course, the applicable values for Hurst exponent $H$ and roughness scale factor $s$ have to be known or estimated. For comparison, we state these properties for crushed meteorite fragments from previous work \cite[fig.~11]{Biele} 
using the data by Nagaashi et al. \cite{Nag} of meteorite fragments of type CM2, CV3, LL3.5 and glass beads. We found $s=1.36\pm0.11$  for Murchison meteorite (CM2) fragments, $s=3.0\pm0.2$ for Allende meteorite (CV3) fragments, $s=4.3\pm0.4$ for NWA 539 (LL3.5) fragments, and, tentatively,  $s=0.23\pm0.03$ for the glass beads (GB); this assumes that $H=1$ (exact) which is consistent with the data and common for natural surfaces.

The temperature dependence of $G_e$ is only in $a(T)$ which depends significantly on the combination of minerals across the contact, i.e., their 
dielectric functions in equation (7). Dislike minerals generally produce 
smaller $a$. For the major rock-forming minerals and other compounds of interest, i.e.,  pyroxenes, 
feldspars, phyllosilicates, olivines, ice Ih, quartz, calcite, dolomite, 
magnesite, silica and industrial glasses, dielectric functions are readily 
available in the literature (usually given as complex refractive index for the
near- and mid-IR) and the integral (7) is easy to evaluate. In a typical rock (fragment), the constituting minerals (with volume fractions assumed to be known) are spatially distributed as grains, lamelleae etc. on many length-scales. It will be the subject of future work to find a proper weighted average of the $a(T)$-factors of every possible combination.

\vskip 0.3cm
{\bf Appendix B}

Let $F(t)$ be the pulsating force exerted on the solid {\bf 1} from an asperity
on solid {\bf 0}. We assume the force localized to the 
point ${\bf x=0}$ on the surface of solid {\bf 1}. Thus the stress
$$\sigma_z ({\bf x},t) = F(t) \delta ({\bf x})$$
act on the solid {\bf 1}. We get
$$\sigma_z ({\bf q},\omega) = {1\over (2\pi )^3} \int d^2x dt \ \sigma_z({\bf x},t)
e^{-i({\bf q}\cdot {\bf x}-\omega t)}$$
$$ = {1\over (2\pi )^2} F(\omega)\eqno(B1)$$
The energy transfer to solid {\bf 1} from the pulsating force $F(t)$ is
$$\Delta E = (2\pi )^3 \int d^2q d\omega \ (-i\omega) u_z({\bf q},\omega) \sigma_z (-{\bf q},-\omega )$$
Using that\cite{Persson}
$$u_z({\bf q},\omega) = M_{zz}(q,\omega) \sigma_z({\bf q},\omega) $$
we get
$$\Delta E = (2\pi )^3 \int d^2q d\omega \ (-i\omega) M_{zz}(q,\omega) |\sigma_z ({\bf q},\omega )|^2$$
Using (B1) this gives
$$\Delta E = {1 \over 2\pi} \int d^2q d\omega \ (-i\omega) M_{zz}(q,\omega) |F(\omega )|^2 $$ 
$$=\int dq d\omega \ q (-i\omega) M_{zz}(q,\omega) |F(\omega )|^2$$
Next if we write $q=(\omega/ c_T) \xi$ (see Ref. \cite{Persson}) then
$$M_{zz} (q,\omega) = f(\xi){c_T^2 \over \omega}$$ giving
$$\Delta E= \int d\xi d\omega \ \xi (-i\omega^2) f(\xi) |F(\omega )|^2$$
Since $\Delta E$ is real we can write
$$\Delta E=  \int_0^\infty d\xi \ \xi {\rm Im} f(\xi) \int d\omega \ \omega^2 |F(\omega )|^2$$
Next note that
$$\int dt \ [\dot F(t)]^2 = $$
$$\int dt d\omega d\omega' (i\omega) (-i\omega')
F(\omega) F(\omega') e^{i(\omega-\omega')t}$$
$$ = 2 \pi \int d\omega \ \omega^2 |F(\omega)|^2$$
Thus we can write
$$\Delta E= {1\over 2 \pi } \int_0^\infty d\xi \ \xi {\rm Im} f(\xi) \int dt \ [\dot F(t )]^2 \eqno(B2)$$
We assume the fluctuation in the position of the atoms are small so that we can expand the
interaction potential $U(d)$ at the asperity contact in the displacement away from the equilibrium separation
$d=d_{\rm eq}$ to linear order in the displacement $s=d-d_{\rm eq}$ so that
$$F(t) = k s$$
where $k=U''(d_{\rm eq})$. Thus we get $\dot F(t) = k v_z(t)$ where $v_z=\dot s(t)$ 
is the velocity of the atom normal to the surface. Thus from (B2) we get
$$\Delta E= {1\over 2 \pi } \int_0^\infty d\xi \ \xi {\rm Im} f(\xi) \int dt \ k^2 v_z^2\eqno(B3)$$
We can write
$$\int dt \ v_z^2(t) = t_0 \langle v_z^2 \rangle$$
where $t_0$ is the time period for which we calculate the energy transfer. Thus the energy transfer per unit time
$$\dot Q_0 = {\Delta E \over t_0} = {1\over 2 \pi } \int_0^\infty d\xi \ \xi {\rm Im} f(\xi)  k^2 \langle v_z^2 \rangle$$
where $\langle v_z^2 \rangle$ is the time average (or ensemble average) of the 
fluctuating atom velocity $v_z(t)$.
We first assume high temperatures so that
$$m \langle v_z^2 \rangle = k_{\rm B} T_0$$
Using this in (B3) gives
$$\dot Q_0 = {1\over 2 \pi } \int_0^\infty d\xi \ \xi {\rm Im} f(\xi)  {k^2 \over m} k_{\rm B} T_0$$
A similar expression with $T_0$ replaced by the temperature $T_1$ gives the energy transfer $\dot Q_1$ from solid
${\bf 1}$ to solid ${\bf 0}$. Thus the net energy transfer is
$$\dot Q = \dot Q_0-\dot Q_1 = G_{\rm a} (T_0-T_1)$$
with
$$G_{\rm a} = {1\over 2 \pi } \int_0^\infty d\xi \ \xi {\rm Im} f(\xi)  {k^2 \over m} k_{\rm B} \eqno(B4)$$
Introducing the friction parameter
$$\eta =
{1\over 2 \pi } {k^2 \over m} \int_0^\infty d\xi \ \xi {\rm Im} f(\xi) $$
we get $G_{\rm a} = k_{\rm B} \eta$.
We have shown in Ref. \cite{Ryberg,Persson} that
$$f(\xi) = {\omega \over c_T^2} M_{zz} (\omega \xi, \omega) = {-i \over \rho c_T^3} {p_L \over (1-2\xi^2)^2 +4 \xi^2 p_T p_L}$$
where
$$p_T=\left (1-\xi^2 +i0\right )^{1/2}$$
$$p_L=\left ([c_T/c_L]^2 -\xi^2 +i0 \right)^{1/2}$$
Thus
$$\eta = {k^2 \mu \over m \rho c_T^3} \eqno(B5)$$
$$\mu = {1\over 2 \pi} \int_0^\infty d\xi \ {\rm Re } {\xi p_L(\xi ) \over (1-2\xi^2)^2 +4 \xi^2 p_T (\xi) p_L (\xi)}\eqno(B6)$$

The derivation above assumes high temperatures, 
but a result valid for all temperatures can be obtained as follows:
Note that phonons are harmonic oscillators where on the average half of the total energy is
potential (elastic deformation) energy and half is kinetic energy. Thus the 
total (phononic) energy $E$ in a solid with $N$ atoms is 
$$E= 2 \times N \times 3 \times {1\over 2} m \langle v_z^2\rangle$$
Thus
$$m \langle v_z^2\rangle = {1\over 3} {E\over N}$$
Thus the heat transfer will depend on
$${1\over 3N} \left [E(T+\Delta T)-E(T) \right] \approx {1\over 3N} 
{\partial E \over \partial T} \Delta T = {C_V\over 3N} \Delta T$$
where $C_V$ is the heat capacity. Thus the heat conductance $G_{\rm a} = k_{\rm B}\eta $ 
is valid for all temperatures if we replace $k_{\rm B}$ with $C_V/3N$.
For high temperatures $C_V \approx 3N k_{\rm B}$ and we recover the limit considered above.

The physical picture behind the heat transfer described above is that  the irregular (random)
thermal movement of the atoms in the contact region (here of atomic size) exert pulsating forces
on the opposite solid (or particle) which result in phonon emission. This emission of phonons
occurs in both direction but is stronger from the hotter to the colder solid. It is assumed the
interaction between the two solids in the contact region (as manifested by the spring constant $k$)
is so weak that it does not influence the irregular motion of the atoms. Here it is interesting to
note that the heat capacity can be expressed as an energy fluctuation term 
$$C_V = k_{\rm B} {\langle \left (E-\langle E \rangle \right )^2 \rangle \over (k_{\rm B} T)^2 }\eqno(B7)$$
so the heat transfer conductance is the product of a friction coefficient (or inverse relaxation time)
and an energy fluctuation term.

Using the Debye model, where the phonon dispersion in the bulk is assumed 
linear for all phonon wavenumbers up to a cut-off wavenumber $\omega_{\rm D}$, one gets 
$$C_V = 9Nk_{\rm B} \left ({T\over T_{\rm D}}\right )^3 \int_0^{x_{\rm D}} dx \ {x^4 e^x \over (e^x-1)^2}$$
where $T_{\rm D} = \hbar \omega_{\rm D} /k_{\rm B}$ and $x_{\rm D} = T_{\rm D} / T$.
The heat capacity for silica is well described by the 
Debye model with $T_{\rm D} \approx 364 \ {\rm K}$ (see Ref. \cite{heat}).
For $T=200 \ {\rm K}$ we get $T/T_{\rm D} \approx 0.55$ and for this relative temperature
the heat capacity has already reached $85\%$ of it high temperature value. Hence using the 
high temperature expression (2) for the damping rate is a good approximation.

\begin{figure} [tbp]
\includegraphics [width=0.47\textwidth,angle=0]{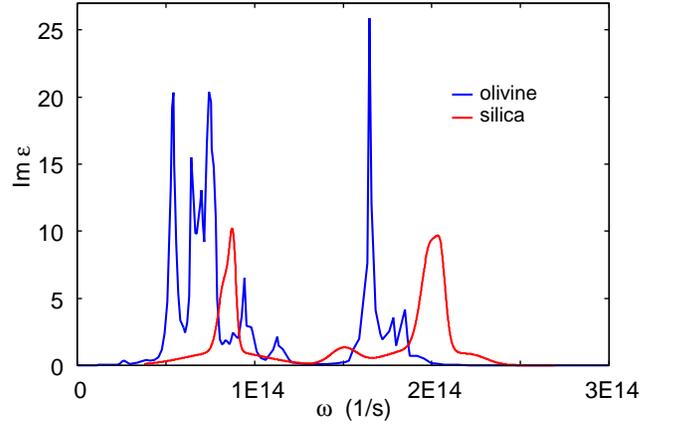}
\caption{
The imaginary part of the dielectric function of silica (red line) and olivine (blue) as a function of frequency in the infrared region.
The loss function structures are due to optical phonons. Data from Ref. \cite{optical} and \cite{YYY}.
}
\label{1Omega.2ImEpsilon.redSilica.blueOlivine.eps}
\end{figure}

\begin{figure} [tbp]
\includegraphics [width=0.47\textwidth,angle=0]{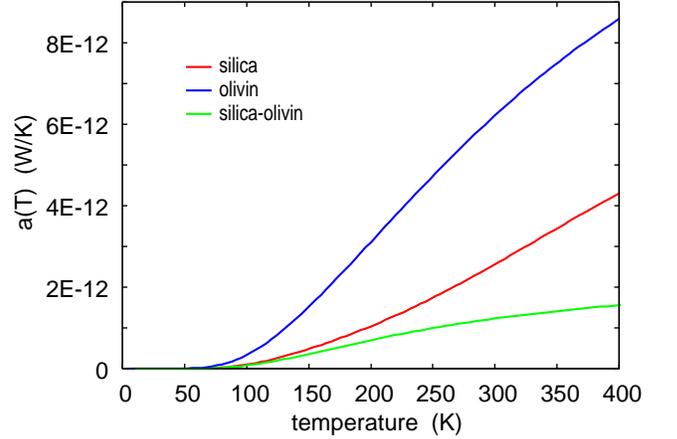}
\caption{
The $a(T)$ factor in the heat transfer coefficient $\alpha_{\rm e} = a/d^2$ as a function of temperature
for two flat silica surfaces (red line), for two olivine surfaces (blue), and for a silica and an olivin surface (green).
}
\label{1Temperature.2aT.SilicaRed.OlivineBlue.eps}
\end{figure}

\vskip 0.3cm
{\bf Appendix C}

In Sec. 4 we have calculated the temperature factor $a(T)$ for silica.
In this appendix we present similar results for another mineral (Mg-rich olivine, 
${\rm Mg}_{1.9}{\rm Fe}_{0.1}{\rm SiO}_4$ ) of interest in applications to
asteroids. However, we note that many minerals occurring in asteroids are inhomogeneous 
%(see Fig. \ref{BASALT.eps} for basalt) 
and in those cases the dielectric function
reflect an average over the different grains in these materials. Since the contact between two mineral
fragments occur in nanometer sized region, and since the contribution to the heat conductance from the
evanescent EM waves is located mainly to a narrow region around the area of real contact, it may be
that the heat transfer involves regions with grains of different chemical composition on the two surfaces.
In that case the two reflection factors $R_0(\omega)$ and $R_1(\omega)$ in (7) will differ, which 
will reduce $a(T)$ as the heat transfer depends on the product ${\rm Im}R_0(\omega ){\rm Im}R_1(\omega)$, 
which depends on ${\rm Im}\epsilon_0 (\omega) {\rm Im}\epsilon_1 (\omega)$.
% Here we will present numerical results where we neglect this different grain effect. 

Fig. \ref{1Omega.2ImEpsilon.redSilica.blueOlivine.eps} shows 
the imaginary part of the dielectric function of silica (red line) and olivine (blue) as a function
of the frequency in the infrared region. In all cases the loss function ${\rm Im}\epsilon(\omega)$ 
is due to optical phonons.
Fig. \ref{1Temperature.2aT.SilicaRed.OlivineBlue.eps} shows 
the $a(T)$ factor in the heat transfer coefficient $\alpha_{\rm e} = a/d^2$ as a function of temperature
for two flat surfaces (separation $d$) of silica (red line) and olivine (blue), and for a silica and olivin surface
(green).

\end{document}